\documentclass[twocolumn,showpacs,preprintnumbers,amsmath,amssymb,prl]{revtex4}
\usepackage{graphicx}
\usepackage{dcolumn}
\usepackage{bm}
\usepackage{amssymb}
\usepackage{amsmath}
\usepackage{epsf}
\usepackage{amssymb}
\usepackage{amsmath}
\usepackage{epsf}
\usepackage{graphics}
\usepackage{verbatim}
\begin{document}

\title{Sharp Magnetic Field Dependence of the 2D Hall Coefficient
Induced by
Classical Memory Effects }
\author{A.~P.~Dmitriev and V.~Yu.~Kachorovskii}
\affiliation{A.F.~Ioffe Physical-Technical Institute, 26
Polytechnicheskaya str., Saint Petersburg, 194021,
Russia}

\date{\today}
\pacs{ 05.60.+w, 73.40.-c, 73.43.Qt, 73.50.Jt}

\begin{abstract}
{ We show that a sharp dependence of the Hall coefficient $R$ on the magnetic field $B$ arises in
two-dimensional electron systems with randomly located strong  scatterers.
The phenomenon is due to classical memory effects. We calculate analytically  the dependence $R(B)$ for the
case of  scattering by hard disks of radius $a$, randomly distributed with concentration $n_0\ll1/a^2$. We
demonstrate that in  very weak magnetic fields ($\omega_c\tau   \lesssim n_0a^2$) memory effects lead to a
considerable renormalization of the Boltzmann value of the Hall coefficient: $\delta R / R \sim 1 .$ With
increasing magnetic field, the relative correction to $R$ decreases, then changes sign, and saturates at the
value $\delta R / R \sim -n_0a^2 .$ We also discuss the effect of the smooth disorder on the dependence of $R$
on $B$. }
\end{abstract}
\maketitle The simplest theoretical description of the magnetotransport properties  of the two-dimensional
(2D) degenerated electron gas is based on the Boltzmann equation which yields the well-known expressions for
the components of the resistivity tensor:
\begin{equation}
\rho_{xx} =  \frac{m}{e^2n \tau}, ~ \rho_{xy} = \frac{m \omega_c} {e^2 n}=- R B. \label{magneto1}
\end{equation}
Here $\tau$ is the transport scattering time, $\omega_c=|e|B/mc$ is the cyclotron frequency, $R= 1/enc < 0$ is
the Hall coefficient, and $n$ is the electron concentration.
Thus, in the frame of the Boltzmann approach, $\rho_{xx}$ and $R$ do not depend on
magnetic field $B$. Experimental measurements of $\rho_{xx}$ and $R$  are widely used to find $\tau$ and $n$.

 It is known,  that Eqs.~\eqref{magneto1} may become invalid
  due to a number of  effects
 of both quantum
 and   classical nature.
 The most remarkable of them
  is the Quantum Hall Effect. Another quantum
 effect, weak localization,  leads to negative magnetoresistance (MR) --   the decrease of $\rho_{xx}$ with
 $B$, concentrated  in the region of weak magnetic fields \cite{Alt}. Besides,  the dependence of $\rho_{xx}$
 on $B$ appears due to
 quantum effects related to electron-electron interaction
 \cite{ee} (see also \cite{gor-ee} for
 review).   At the same time,  both weak localization and electron-electron interaction
 (in frame of standard Altshuler-Aronov
 theory)
 do not result in any dependence of  $R$ on  $B$ (see \cite{fuk} and \cite{ee,zala-ee,gor-ee1}, respectively).

 The dependence of $\rho_{xx}$ on $B$  may also be caused by   classical effects.
One of the reasons is that in
the Boltzmann approach one neglects classical memory effects (ME) arising as a manifestation of non-Markovian
nature of electron dynamics in a static random potential. Physically, a diffusive electron returning to a
certain region of space "remembers"$~$the random potential landscape   in this region, so its motion is not
purely chaotic as it is assumed in the  Boltzmann picture. For $B=0,$ non-Markovian corrections to kinetic
coefficients are usually small. In particular, in a case of hard-core scatterers of radius $a$ (impenetrable
disks) randomly distributed with concentration $n_0,$ ME-induced relative
  correction to the resistivity
 is  proportional to the gas parameter $\beta_0=a/l=2n_0a^2  \ll 1$
 ($l=1/2an_0$ is the mean free path).
  However, for $B\neq 0$
 the role of  ME is dramatically  increased
  due to a  strong dependence of  return probability   on $B.$   Recent studies demonstrated that
 ME lead to a variety  of non-trivial magnetotransport phenomena in 2D disordered systems such as
magnetic-field-induced classical localization \cite{baskin-1,fog-1}, high-field negative
\cite{baskin-1,fog-1,bobylev-1,b4-1,curc0-1} and positive  MR \cite{mir1-1}, low-field anomalous MR
\cite{an1-1,an2-1,k1}, and non-Lorentzian shape of cyclotron resonance \cite{gor-1}.

 In spite of large number of publications, devoted to the   study of the influence  of the non-Markovian
effects on the MR, the dependence of $R$ on $B$,  induced by such effects,  was    investigated (to the best
of our knowledge) only in the context of so-called "circling electrons"$~$\cite{baskin-1}. These electrons
occupy closed cyclotron orbits which avoid scatterers. As a consequence,  they  do not participate in
diffusion. Though the existence of circling  orbits leads to a strong dependence of $\rho_{xx}$ on $B$ in the
region of classically strong $B$ ($\omega_c\tau \gg 1$),  the corresponding dependence of $R$ on $B$  was
found to be very weak in the whole range of $B$ \cite{baskin-1}.

 In this paper, we propose another mechanism of  dependence of $R$   on $B$.
It
  does not rely upon
the existence of non-colliding electrons but, in contrast, assumes that transport properties of colliding
electrons are
modified by  classical  ME. The mechanism turns out to be  especially  effective in the region of very weak
fields,  $\omega_c \tau \lesssim \beta_0.$

We will study dependence of $R$ on $B$ in 2D degenerated electron gas  in a system of randomly located hard core
scatterers. We restrict ourselves to the study of the case of classically weak fields ($\omega_c\tau
\ll 1$) and assume that Fermi wavelength $\hbar/mv_F$ is much smaller than $a.$ The latter assumption will
allow us to study the electron dynamics on the classical level.

We start with recalling that in the frame of the  Boltzmann approach, the collision with a single scatterer is
described by differential scattering cross-section $\sigma(\theta)$ (see Fig.~1a) and  the collisions with
different scatterers are independent. Inverting in time the process shown in Fig.~1a we get a  process shown
in Fig.~1a$^{\prime}$, corresponding to scattering by the angle $-\theta$. This implies an important property
of a single scattering -- the symmetry with respect to replacement of $\theta$ by $-\theta$ (reciprocity
theorem): $\sigma(\theta) = \sigma(-\theta)$ \cite{kom}.
This is the property which provides that $R$ does not depend on $B$. If, for any reason,  scattering
cross-section acquires an asymmetric correction $\delta \sigma(\theta) \neq \delta \sigma(-\theta)$, the
expression for $\rho_{xy}$  becomes $\rho_{xy} = {m }(\omega_c +\Omega) /{e^2 n}= - B(R+\delta R),$ where
\begin{equation}
\Omega= - n_0v_F\int
d\theta\delta \sigma(\theta)\sin\theta,~~~\frac{\delta R}{R}=\frac{\Omega}{\omega_c},
 \label{Omega}
\end{equation}
and $v_F$ is the Fermi velocity.
 In particular, such an asymmetric correction arises due to
 ME specific for processes of double
scattering on a scatterer  after return to it (see Fig.~1b,b$^\prime$,c,c$^\prime$). Though such processes are
beyond the Boltzmann picture, they can be  formally included into the kinetic equation by a slight
modification of the Boltzmann collision integral. Specifically, one can introduce a small change of the
scattering cross-section $\sigma(\theta) \to \sigma(\theta)+\delta \sigma(\theta)$ on the disk where double
scattering takes place (disk 1  in Fig.~1b,b$^\prime$,c,c$^\prime$)
 \cite{an2-1,lub}. For $B=0$, cross-section remains symmetric:
$\delta\sigma(\theta)=\delta\sigma(-\theta).$
\begin{figure}[ht!]
 \leavevmode \epsfxsize=6.5cm \centering{\epsfbox{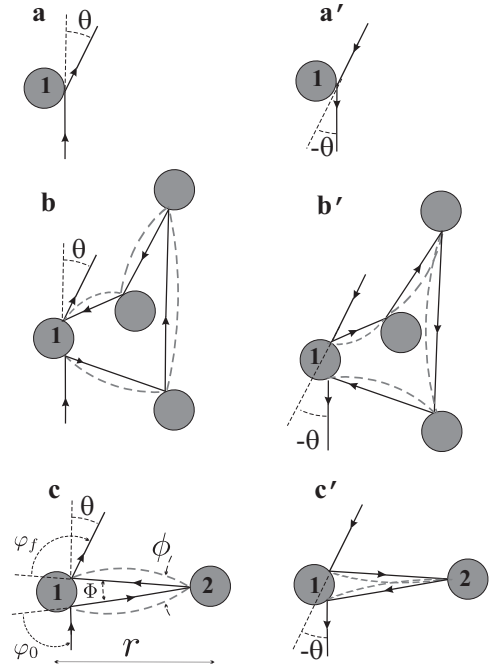}}
\caption{  Processes of single scattering
   by angle $\theta$ (a) and $-\theta$ (a$^\prime$) characterized by a   scattering  cross-section
  $\sigma(\theta)$ ($\sigma(\theta)=\sigma(-\theta)$ both
for  $B=0$ and for  $B \neq 0$),
    and processes of
        scattering on  complexes  of
scatterers (b,b$^\prime$,c,c$^\prime$) including double scattering on scatterer  1.  Correction to the
cross-section due to multi-scattering processes remains symmetric for $B=0.$ Magnetic field bends trajectories
as shown in b,b$^\prime$,c,c$^\prime$ by dashed lines. As a result, the symmetry with respect to inversion of
$\theta$ is broken, so that $\delta\sigma(\theta)\neq \delta\sigma(-\theta)$  for $B\neq 0.$} 
\end{figure}
 However, for $B \neq 0$
the time inversion symmetry is broken, so that the  cross-section becomes asymmetric: $\delta \sigma(\theta)
\neq \delta \sigma(-\theta).$  The point is that the influence of the magnetic field is different for the
processes where closed return path is passed counterclockwise (Fig.~1b,c) and clockwise
(Fig.~1b$^\prime$,c$^\prime$).

The  return after one scattering (see Fig.~1c,c$^\prime$) needs special attention because the probability of
such a process very sharply depends on $B$ due to  "empty corridor effect" $~$\cite{an1-1,an2-1}. The physics
behind this effect is as follows \cite{an1-1}.  The passage of an electron from disk $1$ to disk $2$ ensures
the absence of the disk's centers in the region of width $2a$ around this part of trajectory (from $1$ to
$2$). In other words, there exists an empty corridor with an area $S= 2a r,$ (here $r$ is the distance between
disks) surrounding the segment $1 \rightarrow  2$ of the electron trajectory. This reduces the scattering
probability on the way back. The total probability $W=W_1W_2$ of the passage $1 \rightarrow 2 \rightarrow 1$
is the product of the probability $W_1=\exp(-n_0S)=\exp(-r/l)$ of passage form $1$ to $2$ and the probability
$W_2=\exp(-n_0[S-S_0]) = \exp(-r/l +n_0 S_0)$ of the passage $2 \rightarrow 1.$ Here
 $S_0$
  is the area of the overlap of the two corridors, surrounding segments $1\rightarrow 2$ and $2 \rightarrow 1$,
  respectively. Hence,
 $W=\exp\left( -2r/l+n_0 S_0 \right).$
 The magnetic field  pulls out (together) forward and backward trajectories for process shown in Fig.~1c (Fig.~1c$^\prime$)
thus decreasing (increasing) $S_0$ and leading to a sharp dependence of $W$ on $B$. The corresponding
correction to $\rho_{xx}$ was calculated numerically in \cite{an1-1} and  analytically in \cite{an2-1}.

   The calculation of  $R$  is quite analogous to the calculation of $\rho_{xx}$  presented in \cite{an2-1}.
      The easiest way to find $R$ is to use the
    expression for $\delta \sigma (\theta)$
    derived  in \cite{an2-1}:
\begin{align}
&\delta\sigma(\theta)=\frac {1}{4l}  \int_a^{\infty} \frac{dr}{r}e^{-2r/l} \int_0^{2\pi} d \varphi_0
\int_0^{2\pi} d \varphi_f \nonumber
\\
&   \sigma(\varphi_0)\sigma(\varphi_f) e^{ n_0S_0 }[\delta(\theta-\varphi_{\varphi_0,\varphi_f})+
 \delta(\theta-\pi)
 \nonumber
\\
&-\delta(\theta-\varphi_{\varphi_0,0})-\delta(\theta-\varphi_{0,\varphi_f})]. \label{effective1}
 \end{align}
 Here $\varphi_{\varphi_0,\varphi_f}= (\pi +\varphi_0+\varphi_f)({\rm mod}~2\pi ),$
    $\sigma(\varphi)=(a/2)|\sin(\varphi/2)|$ is the single scattering cross-section,
  $S_0
 = \int_0^r d r^{\prime}
\left(2a - \left\vert \phi r' - r'^2/R_c \right\vert\right) \theta \left[2a - \left\vert \phi r' - r'^2/R_c
\right\vert \right] ,$
 $\theta[\cdots]$ is the Heaviside step function,
$\phi =\Phi +r/R_c,$ $R_c$ is the cyclotron radius and $\Phi \approx ({a}/{r})[\cos(\varphi_0/2) +
\cos(\varphi_f/2)]$ (see Fig.~1c). Introducing dimensionless variables $T=r/l, \ z= \omega_c\tau/\beta_0$ and
using Eq.~\eqref{Omega}, we get
\begin{align}
&\frac{\delta R}{R}=g(z)= -  \int_0^{\infty}\frac{dT}{T} e^{-2T} \int_0^{\pi} d \alpha \int_0^{\pi} d \gamma
\nonumber
\\ & \times \sin\left(\alpha+\gamma\right) \sin^2\alpha \sin^2 \gamma ~\frac{ e^{ s_z} - e^{ s_0}}{2z}.
\label{g(z)} \end{align} Here $s_z= \int_0^{T} dt \left(1-\left\vert \zeta t -\frac{z t^2}{2} \right\vert
\right) \theta \left(1- \left\vert \zeta t -\frac{z t^2}{2} \right\vert \right),$ $\zeta={(\cos\alpha
+\cos\gamma)}/{2T} +{z T}/{2}, \quad s_0= s_{z \to 0}.$ Function $g(z)$ calculated numerically with the use of
Eq.~\eqref{g(z)} is plotted in Fig.~2. For    $z\ll 1 $, $g(z)\approx 0.064 - 4 z^2.$ For  $z \gg 1$, $g(z)$
decreases as $0.35/z^{3/2}$. It worth emphasizing that  $\delta R/R \sim 1$ for $z \lesssim 1.$  This means that
the correction is not parametrically small in a gas parameter $\beta_0$ which is usually considered as
expansion parameter for  ME-induced corrections.
\begin{figure}[ht!]
\vspace{-3mm}
 \leavevmode \epsfxsize=7cm \centering{\epsfbox{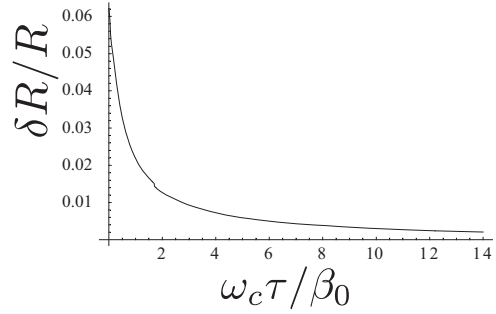}} %
 \vspace{-3mm}
\caption{Magnetic field dependence of the relative correction to the Hall coefficient caused by empty corridor effect.
 \vspace{-3mm}
 }
\end{figure}

Next we calculate   $\delta R$ for stronger fields, $\beta_0 \ll \omega_c\tau \ll 1.$ At such fields empty
corridor effect is suppressed and  returns after one scattering (Fig.~1c,c$^\prime$) and after a number of
scatterings  (Fig.~1b,b$^\prime$)   equally contribute to $\delta R.$ In this case, one can also introduce
the effective scattering cross-section \cite{lub} which turns out to be frequency-dependent and for $\omega =
0$  reads \cite{kom1}
\begin{align}
&\delta \sigma(\theta-\theta')= v_F \int [\sigma(\theta-\varphi)-\sigma_0\delta(\theta-\varphi)]
\label{sig}  \\
& \times \tilde G(0, \varphi-\varphi') [\sigma(\varphi'-\theta')-\sigma_0\delta(\varphi'-\theta')]d\varphi d\varphi'.
\nonumber
\end{align}
Here $\sigma_0=\int d\varphi\sigma(\varphi) $ is the total cross-section for single scattering, $\tilde G(0,
\varphi-\varphi')=\tilde G(\mathbf r, \varphi,\varphi')|_{\mathbf r\to 0}, ~\tilde G(\mathbf r,
\varphi,\varphi')=
 G(\mathbf r, \varphi,\varphi')- G^{\rm ball}(\mathbf r, \varphi,\varphi'),$
 $G(\mathbf r, \varphi,\varphi')$ is the Green function of the stationary Boltzmann equation,
$\displaystyle
 G^{\rm ball}(\mathbf r, \varphi,  \varphi^\prime ) =\frac{ \exp\left(-\theta_r/\beta\right)
}{v_Fr\cos(\theta_r/2)}
\delta( \varphi - \varphi_\mathbf r+\theta_r/2) \delta(\varphi^\prime - \varphi_\mathbf r-\theta_r/2)
$
is  the Green function of the Boltzmann equation without
 in-scattering term, $\varphi_\mathbf r$ is the angle of vector $\mathbf r,$
  and  $\theta_r=2\arcsin(\beta r/2l).$
  Substituting Eq.~\eqref{sig} into Eq.~\eqref{Omega} and using the property   $ \int d\varphi
d\varphi_0 G(0,\varphi,\varphi^\prime)\sin(\varphi-\varphi^\prime)=0$ \cite{bbb}, we get after some algebra
\begin{equation}
{\delta R}/{R} = -  {n_0\sigma_{\rm tr}^2}/{2\pi} \ll 1, \label{hall-fin}
\end{equation}
where $\sigma_{\rm tr}=\int d\theta\sigma(\theta)(1-\cos\theta)=8a/3.$ Hence, with increasing $B$ relative
correction decreases according to Eq.~\eqref{g(z)}, then changes sign and saturates at small negative value.
It is noteworthy that, as follows from the above derivation, Eq.~\eqref{hall-fin} is valid not
only for the case of impenetrable disks but also for any type of well-separated scatterers.

Above we discussed an idealized system where only strong scatterers are present. Let us now assume that in
addition to strong scatterers there is
   a  weak smooth random potential $U(\mathbf r)$ with the  rms amplitude $U$  and the correlation
length $d$ ($a \ll d \ll l$). The presence of such a potential does not influence  the empty corridor effect
provided that $\lambda  \gg l,$ where $\lambda \sim d (E_F/U)^{2/3}$ is the Lyapunov length, characterizing
the divergence of the electron trajectories in the  potential $U(\mathbf r).$ In the opposite limit, $\lambda
\ll l$, one should restrict integration over $r$ in Eq.~\eqref{effective1} by $\lambda.$   In this case,
relative correction to $R$ decreases: $\delta R/R \sim \lambda/l.$  On the other hand, the field needed for
suppression of the  empty corridor effect increases and can be found from the following estimate $\omega_c\tau
\sim \beta_0 (l/\lambda)^2.$ At such a field two corridors, corresponding to passage $1\rightarrow 2$ and $2
\rightarrow 1$ (see Fig.~1c,c$'$) between  disks $1$ and $2$ separated by a distance $r \approx \lambda,$
cease to overlap.

To study the effect of the smooth disorder at stronger fields, one should add a term $(\mathbf
F/m)\partial/\partial \mathbf v$ in the l.h.s.  of the Boltzmann equation, where $\mathbf F=-\partial
U/\partial \mathbf r$. Treating this term as a small perturbation, we find correction to the Boltzmann
collision integral $\displaystyle \delta \hat T =   \langle
 \mathbf F(\partial/\partial \mathbf v)~
 \hat G~
 \mathbf F(\partial/\partial \mathbf v)/m^2
 \rangle,  $ where $\langle\cdots \rangle$ stands for averaging over  realizations of $U(\mathbf r)$ and
$\hat G$ is the operator with the kernel $G(\mathbf r, \varphi,\varphi^\prime)$. This kernel  can not be found
explicitly for the above-discussed  case of scattering on impenetrable disks.  However, the exact solution is
possible for short range scatterers, where $\sigma(\theta)=\sigma_0/2\pi=const$. In this case, after
cumbersome but straightforward calculations we get (in addition to Eq.~\eqref{hall-fin})
\begin{equation}
\hspace{-1.7mm}
 \frac{\delta R}{R} \approx -\frac{17(\omega_c\tau)^2}{192 E_F^2 l^2}\int_0^{\infty}dr r\kappa(r) \sim -
(\omega_c\tau)^2 \frac{d^2}{l^2}\frac{U^2}{E_F^2}.
\nonumber
\end{equation}
We see that smooth disorder leads to appearing of a very weak parabolic dependence of $R$ on $B$.

To conclude, we have shown that in a 2D system with rare hard-core scatterers classical ME lead to a very
sharp dependence of $R$ on $B$ concentrated in the region of very weak fields ($\omega_c\tau \lesssim a/l$).
The total variation of $R$ in this region of fields is on the order of the Boltzmann value of $R.$ At larger
fields, where $a/l \ll \omega_c\tau \ll 1 ,$ the ME lead to a small field-independent correction to $R$ and
(in a presence of smooth disorder) to a very weak parabolic dependence.

We acknowledge I.~Gornyi and G.~Minkov for fruitful discussion. The work was  supported by RFBR, by
grant of Russian Scientific School, and by programmes of the RAS.
 V.Yu.K. was  supported by  Dynasty foundation.


\begin{thebibliography}{8}
\bibitem{gor-ee1} In fact, a smooth   dependence of $R$ on $B$ may
arise due to interaction-induced  correction to the density of states in the Cooper channel, I.~Gornyi,
privite communication (see also \cite{ee}).
\bibitem{kom} Here we neglect the effect of magnetic field on a single scattering event.
\bibitem{kom1}
Eq.~\eqref{sig} is obtained by integration over time  Eq.~(13) of \cite{lub} end extracting contribution of ballistic term. The latter one   represents  propagation without collision and, evidently, can not give any contribution to return processes.
\bibitem{bbb} To obtain this property we first integrate the Boltzmann equation (in $q-$space) over $\varphi$ which yields
$\int G_{\mathbf q}(\varphi,\varphi^\prime)i\mathbf q \mathbf v d\varphi =1.$ One can show that integration
over $\varphi^\prime$ yields the same result $\int G_{\mathbf q}(\varphi,\varphi^\prime)i\mathbf q \mathbf
v^\prime d\varphi^\prime= 1.$ Using these identities we get $ \int d\varphi d\varphi^\prime d \mathbf q
G_{\mathbf q}(\varphi,\varphi^\prime)\sin(\varphi-\varphi^\prime)=\int d\varphi d\varphi^\prime d \mathbf q
G_{\mathbf q}(\varphi,\varphi^\prime)[\sin(\varphi-\varphi_\mathbf q)\cos(\varphi_{\mathbf q}-\varphi^\prime)+
\cos(\varphi-\varphi_\mathbf q)\sin(\varphi_{\mathbf q}-\varphi^\prime)]=0.$
\end{thebibliography}
\end{document}